%% file: Near-Field_Spatial_Correlation_for_Extremely_Large-Scale_Array_Communications.tex
\documentclass[journal]{IEEEtran}

\usepackage{cite}
\usepackage{graphicx}
\usepackage{subfigure}
\usepackage{url}
\usepackage{epsfig}
\usepackage{amsmath}
\usepackage{amsfonts}
\usepackage{amssymb}
\usepackage{mathrsfs}
\usepackage[final,scrtime]{prelim2e}  
\usepackage{verbatim}
\usepackage{array}
\usepackage{algorithm}
\usepackage{algorithmic}
\usepackage{amsthm}
\usepackage{color}
\usepackage{lipsum,mwe,cuted}
\usepackage{stfloats}

\input{./Definitions.tex}

\newcommand{\bibdir}{../Bibtex}
\begin{document}

\title{Near-Field Spatial Correlation for Extremely Large-Scale Array Communications}

\author{Zhenjun Dong and Yong Zeng,~\IEEEmembership{Member,~IEEE}

       \thanks{Z. Dong and Y. Zeng are with the National Mobile Communications Research Laboratory, Southeast University, Nanjing 210096, China. Y. Zeng is also with the Purple Mountain Laboratories, Nanjing 211111, China (e-mail: \{zhenjun\_dong, yong\_zeng\}@seu.edu.cn). (\emph{Corresponding author: Yong Zeng.})
}

}

\maketitle
\begin{abstract}
Extremely large-scale array (XL-array) communications correspond to systems whose antenna sizes are so large that the scatterers and/or users may no longer be located in the far-field region.
By discarding the conventional far-field uniform plane wave (UPW) assumption, this letter studies the near-field spatial correlation of XL-array communications, by taking into account
the more generic non-uniform spherical wave (NUSW) characteristics. 
It is revealed that different from the far-field channel spatial correlation which only
depends on the \emph{power angular spectrum} (PAS), the near-field spatial correlation
depends on the scattered power distribution not just characterized by their arriving
angles, but also by the scatterers' distances, which is termed as \emph{power location spectrum} (PLS).
A novel integral expression is derived for the near-field spatial correlation in terms of
 the scatterers' location distribution, which includes the far-field spatial correlation
 as  a  special case. The result shows that different from the far-field case,
 the near-field spatial correlation no longer exhibits \emph{spatial stationarity} in general,
 since the correlation coefficient for each pair of antennas depends on their specific positions, rather than their relative distance only.
To gain further insights,
 we propose a generalized one-ring model for scatterer distribution,
 by allowing the ring center to be flexibly located rather than coinciding with the
 array center as in the conventional one-ring model.
Numerical results are provided to show the necessity of the near-field spatial correlation modelling for XL-array communications.

\end{abstract}

\section{Introduction}
Massive multiple-input multiple-output (MIMO) has become a reality in the fifth-generation (5G) mobile communication networks, with 64 antennas typically deployed at the base station~\cite{bjornson2019massive}.
To further improve the spatial resolution
and communication spectral efficiency for beyond 5G (B5G) networks,  there has been growing interest in further increasing  the antenna size/number drastically,
leading to communication systems known as ultra-massive MIMO~\cite{akyildiz2016realizing}, extremely large aperture massive MIMO~\cite{amiri2018extremely}, or extremely large-scale MIMO (XL-MIMO)~\cite{de2020non}.

As the antenna size goes large, it is more likely that the users and/or scatterers are located in the near field of the extremely large-scale array (XL-array)~\cite{lu2021does}, i.e., their distance is smaller than the Rayleigh distance that increases quadratically with the array dimension~\cite{balanis2015antenna}.
In this case, the commonly adopted uniform plane wave (UPW) assumption no longer holds.
Instead, the more generic non-uniform spherical wave (NUSW) characteristics need to be taken into account to accurately model the power and phase relationships across different array elements.
Some preliminary research efforts have been made towards this direction~\cite{lu2021does,9617121,bjornson2020power}.  
For example, a unified near-field modelling is developed in~\cite{9617121} for extremely large-scale discrete array and continuous surface,
based on which a closed-form expression of the maximal signal-to-noise ratio (SNR) is derived in terms of the geometric relationship formed by the XL-array/surface and the user location.
In~\cite{bjornson2020power}, a closed-form channel gain expression is derived by considering a planar array of arbitrary size, which captures essential near-field behaviors and revisits the power scaling laws.

However, the aforementioned existing works on the near-field modelling and performance analysis are based on the assumption of free-space line-of-sight (LoS) propagation.
For most practical communication scenarios, signals are subject to multi-path fading due to the random scattering, reflection, and diffraction.
In this case, channels are typically modelled stochastically and the spatial correlation is of paramount importance for the second-order statistical channel characterization.
For instance, accurately characterizing the spatial correlation is necessary for developing the Kronecker channel model~\cite{forenza2006benefit},
as well as for deriving the optimal transmission strategy based on stochastic channel state information~\cite{jorswieck2004channel}.
The spatial correlation based on the far-field UPW assumption has been extensively studied~\cite{abdi2002space,4277097}, which depends on the scattered power distribution characterized by \emph{power angular
spectrum} (PAS) and exhibits spatial stationarity, i.e., the channel correlation coefficient between each pair of array elements only depends on their relative distance along the array~\cite{abdi2002space}.
However, such results cannot be applied for the near-field spatial correlation for XL-array communications.

This letter focuses on the near-field modelling and spatial correlation characterization for the basic single-input multiple-output
(SIMO) communication with XL-array. By taking into account the generic NUSW characteristics, a novel integral expression is derived for the near-field spatial correlation,
which includes the conventional far-field spatial correlation as a special case.
It is revealed that the near-field spatial correlation in general depends on the scattered power distribution not just characterized by their
arriving angles as in UPW model, but also by the scatterers' distances from the array, which we term as \emph{power location spectrum} (PLS).
Furthermore, the near-field spatial correlation coefficient no longer exhibits spatial stationarity, since the correlation for each pair of array elements depends on their
actual positions along the array, rather than their relative distance only.
To gain further insights, we propose a generalized one-ring model
for the scatterer distribution, by allowing the ring center to be
flexibly located rather than coinciding with the array center
as in the conventional one-ring model~\cite{837052}.
Numerical results demonstrate the necessity of the near-field spatial correlation modelling, since it leads to quite a different result than the conventional far-field modelling
in terms of the eigenvalue distribution of the correlation matrix.

\section{System Model}
\begin{figure}
  \centering
  \includegraphics[width=0.9\columnwidth]{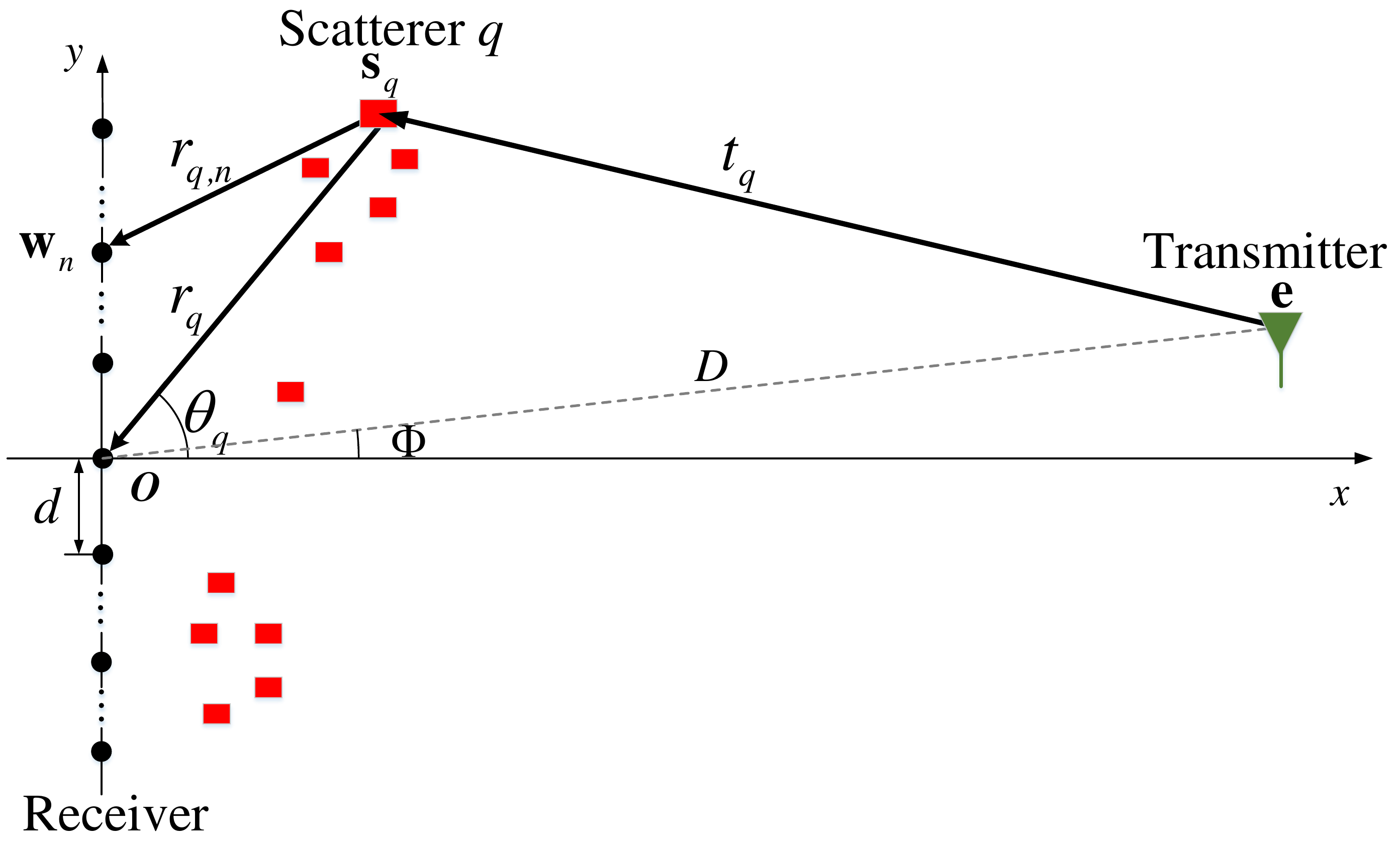}
  \caption{Extremely large-scale array communication in near-field scattering environment.}
  \label{fig:Spherical wave1}
\end{figure}

As shown in Fig.\ref{fig:Spherical wave1}, we consider an XL-array communication system,  where a single-antenna transmitter communicates with a receiver that is equipped with an XL-array of $N\gg 1$ elements.
We focus on the basic ULA architecture with adjacent elements separated by distance $d$. Without loss of generality, we assume that the ULA is placed along the $y$-axis of the Cartesian coordinate system
, and the $n$th array element is located at $\vec{w}_n=[0,nd]^T$, where $-\left\lceil \frac{N-1}{2}\right\rceil\leq n\leq \left\lfloor\frac{N-1}{2}\right\rfloor$, with $\lceil\cdot\rceil$ and $\lfloor\cdot\rfloor$ denoting the ceiling and floor operations, respectively.
The location of the transmitter is denoted by $\vec{e}=[D\cos\Phi,D\sin\Phi]^T$,
where $D$ denotes the distance between the transmitter and the origin,
and $\Phi\in\left[-\frac{\pi}{2},\frac{\pi}{2}\right]$ is the angle relative to the positive $x$-axis.
For scattering environment with $Q$ scatterers,
denote the location of scatterer $q$ as $\vec{s}_q=[r_{q}\cos{\theta_q},r_{q}\sin{\theta_q}]^T$, $1\leq q\leq Q$,
where $r_{q}$ and $\theta_q\in\left[-\pi,\pi\right]$ denote the distance from the origin and the angle relative to positive $x$-axis, respectively.
Hence, the distance between scatterer $q$ and the $n$th array element is
\begin{equation}\label{eq:r_q}
r_{q,n}=\|\vec{s}_q-\vec{w}_n\|=r_{q}\sqrt{1+\left(nd/r_q\right)^2-2\sin{\theta_q}nd/r_q},
\end{equation}
where $r_{q,0}=r_{q}$.
Similarly, the distance between the transmitter and scatterer $q$ is
\begin{equation}\label{eq:t_q}
  t_{q}=\|\vec{s}_q-\vec{e}\|=\sqrt{r_{q}^2+D^2-2r_{q}D\cos(\theta_q-\Phi)}.
\end{equation}

Note that most of the existing works assume that the scatterers are located
in the far-field of the antenna array, so that UPW modelling is adopted for
each scattering path~\cite{balanis2015antenna}. Such an assumption becomes
invalid in the XL-array regime when the scatterers are located in the near-field region.
Therefore, we aim to study the channel modelling and spatial correlation of XL-array
communication with the more generic NUSW model.
In this case, the exact distances in (1), rather than its first-order Taylor approximation needs to be used to accurately model the amplitude and phase relationships across array elements.
To that end, based on the bistatic radar equation due to scattered rays~\cite{skolnik1980introduction},
the channel coefficient $h_n$ between the transmitter and the $n$th array element, can be modelled as
\begin{equation}\label{eq:h_rcs}
   {h}_n=\sum_{q=1}^{Q}\frac{\lambda\sqrt{\sigma_{q}}}{(4\pi)^{3/2}t_{q}r_{q,n}}e^{-j\frac{2\pi}{\lambda}(t_{q}+r_{q,n})+j\psi_{q} },
\end{equation}
where $\lambda$ is the signal wavelength, $\sigma_{q}$ represents the radar cross
section (RCS) of scatterer $q$, which is modelled as independent and identically distributed
(i.i.d.) positive random variables for different $q$~\cite{skolnik1980introduction};
$\psi_{q}$ represents the phase shift due to scatterer $q$, which is modelled as i.i.d. random variable with uniform distribution over $[-\pi,\pi)$.
It is not difficult to see that (3) can be equivalently written as
\begin{equation}\label{eq:h_rcs2}
   {h}_n=\sum_{q=1}^{Q}\frac{\lambda\sqrt{\sigma_{q}}}{(4\pi)^{3/2}t_{q}r_{q}}\frac{r_{q}}{r_{q,n}}e^{-j\frac{2\pi}{\lambda}(t_{q}+r_{q,n})+j\psi_{q}}.
\end{equation}

Note that the component $\frac{\lambda\sqrt{\sigma_{q}}}{(4\pi)^{3/2}t_{q}r_{q}}$ in (\ref{eq:h_rcs2}) only
depends on the RCS and location of scatterer $q$, while independent of the array element index $n$. Therefore, we may define
$\frac{\lambda\sqrt{\sigma_{q}}}{(4\pi)^{3/2}t_{q}r_{q}}=\sqrt{\frac{\beta_0}{Q}}g_q$,
where $\beta_0=\mathbb{E}[|h_0|^2]$ is the average received power for the reference array element $n=0$, and
$g_{q}$ is a random variable corresponding to the signal amplitude at array element $n=0$ that is contributed by
scatterer $q$, with $Q^{-1}\sum_{q=1}^{Q}\mathbb{E}[g_{q}^2]=1$. Note that such a definition is motivated by the far-field spatial correlation modelling in~\cite{abdi2002space}, by considering the more generic near-field NUSW model.
As such, the channel in (3) and (4) for NUSW can be equivalently expressed in terms of the average power $\beta_0$ at the reference array element $n=0$ as
\begin{equation}\label{eq:hii}
   {h}_n=\sqrt{\frac{\beta_0}{Q}}\sum_{q=1}^{Q}g_{q}\frac{r_q}{r_{q,n}} e^{-j\frac{2\pi}{\lambda}(t_{q}+r_{q,n})+j\psi_{q} }.
\end{equation}

Let $\vec{h} \in \mathbb{C}^{N\times 1}$ denote the channel vector containing the channel coefficients $h_n$ in (5) for all the $N$ array elements,
and $\vec{r}_q\in \mathbb{R}^{N\times 1}$ contains all the $N$ distances $r_{q,n}$ associated with scatterer $q$. Then (5) can be written in a vector form as
\begin{equation}\label{eq:H_rcs}
   \vec{h}=\sqrt{\frac{\beta_0}{Q}}\sum_{q=1}^{Q}g_qr_qe^{-j\frac{2\pi}{\lambda}t_{q}+j\psi_{q} }\frac{1}{\vec{r}_{q}}\circ e^{-j\frac{2\pi}{\lambda}\vec{r}_q},
\end{equation}
where $\circ$ denotes the Hadamard product operation.
Our objective is to characterize the spatial correlation matrix
$\vec{R}=\mathbb{E}\left[\vec{h}\vec{h}^H\right]$, based on the near-field NUSW model in (6).
\section{near-field spatial correlation}
Based on (\ref{eq:hii}), the near-field spatial correlation between array elements $n$ and $m$ can be expressed as
\begin{equation}\label{eq:Rnm}
\begin{aligned}
&\vec{R}_\text{NF}(n,m)=\mathbb{E}\left[{h}_n{h}_m^*\right]\\
&=\frac{\beta_0}{Q}\sum_{q=1}^{Q}\sum_{p=1}^{Q}\frac{r_qr_pe^{-j\frac{2\pi}{\lambda}(t_{q}+r_{q,n}-t_{p}-r_{p,m})}\mathbb{E}[g_{q}g_{p}]\mathbb{E}\left[e^{j\psi_{q}-j\psi_{p}}\right ]}{r_{q,n}r_{p,m}},
\end{aligned}
\end{equation}
where the subscript $(\cdot)_\text{NF}$ signifies that the spatial correlation is based on the near-field NUSW model.
With $\psi_q$ following i.i.d. uniform distribution over $[-\pi, \pi)$, we have $\mathbb{E}\left[e^{j\psi_{q}-j\psi_{p}}\right]=\delta(p-q)$.
Hence, $\vec{R}_\text{NF}(n,m)$ in (7) can be simplified to
\begin{equation}\label{eq:Rnm}
\vec{R}_\text{NF}(n,m)=\frac{\beta_0}{Q}\sum_{q=1}^{Q}\frac{r_q^2e^{-j\frac{2\pi}{\lambda}(r_{q,n}-r_{q,m})} \mathbb{E}\left[g_q^2\right]}{r_{q,n}r_{q,m}}.
\end{equation}

Similar to the far-field UPW modelling in~\cite{abdi2002space}, for $Q\rightarrow \infty$, $Q^{-1}\mathbb{E}\left[g_q^2\right]$ in (8) can be interpreted as the infinitesimal power contributed by a differential scatterer located around $\vec{s}_q$ to the reference array element $n=0$.
Thus, we may express $Q^{-1}\mathbb{E}\left[g_q^2\right]=f(\vec{s}_q)d\vec{s}$, where $f(\vec{s})$ denotes the probability distribution function (PDF) of the scatterer location $\vec{s}\in \vec{S}$, with $\vec{S}$ denoting the support of random scatterers.
Hence, for $Q\rightarrow\infty$, $\vec{R}_\text{NF}$ in (8) can be expressed in an integral form as
\begin{equation}\label{eq:cov1nm}
\vec{R}_\text{NF}(n,m)=\beta_0\int_{\vec{s}\in\vec{S}} \frac{r^2(\vec{s})}{r_n(\vec{s})r_m(\vec{s})}e^{-j\frac{2\pi }{\lambda}\left(r_n(\vec{s})-r_m(\vec{s})\right)} f(\vec{s})\text{d}\vec{s},
\end{equation}
where $r(\vec{s})=r_0(\vec{s})$ and $r_n(\vec{s})=\|\vec{s}-\vec{w}_n\|$ denotes the distance between the scatterers and the $n$th array element.

It is worth remarking that under the UPW assumption, the far-field spatial correlation can be expressed as~\cite{abdi2002space}
\begin{equation}\label{eq:Rnm_UPW}
\vec{R}_\text{FF}(n,m)=\beta_0\int_{-\pi}^{\pi} e^{-j\frac{2\pi }{\lambda}(m-n)d\sin\theta} f(\theta)\text{d}\theta,\\
\end{equation}
where $\theta$ is the angle of arrival (AoA). A direct comparison between (9) and (10) reveals two important differences between the near- and far-field spatial correlations.
Firstly, while the far-field spatial correlation only depends on the PAS $f(\theta)$, that for the near-field model depends on the scattered power distribution not just characterized by their
arriving angles, but also by the scatterers' distances, for which we term $f(\vec{s})$ in (9)
as the PLS. Secondly, while the far-field correlation is spatially wide-sense stationary, since $\vec{R}_\text{FF}(n,m)$ in (10) only depends on the array index difference $m-n$, such a stationary property does not hold in general for the near-field correlation in (9).

\emph{Lemma 1:} When $Nd\ll r(\vec{s})$, $\forall\vec{s}\in\vec{S}$, we have
\begin{equation}\label{eq:Rnm_UPW2}
\vec{R}_\rm{NF}(n,m)\approx \vec{R}_\text{FF}(n,m).
\end{equation}

\begin{proof}
Please refer to Appendix A.
\end{proof}
Lemma 1 shows that our developed near-field spatial correlation in (\ref{eq:cov1nm}) generalizes the conventional far-field spatial correlation in (\ref{eq:Rnm_UPW}), since the latter is included as a special case
when the scatterers are in the far field of XL-array.

\section{Generalized One-Ring model}
\begin{figure}
  \centering
  \includegraphics[width=0.89\columnwidth]{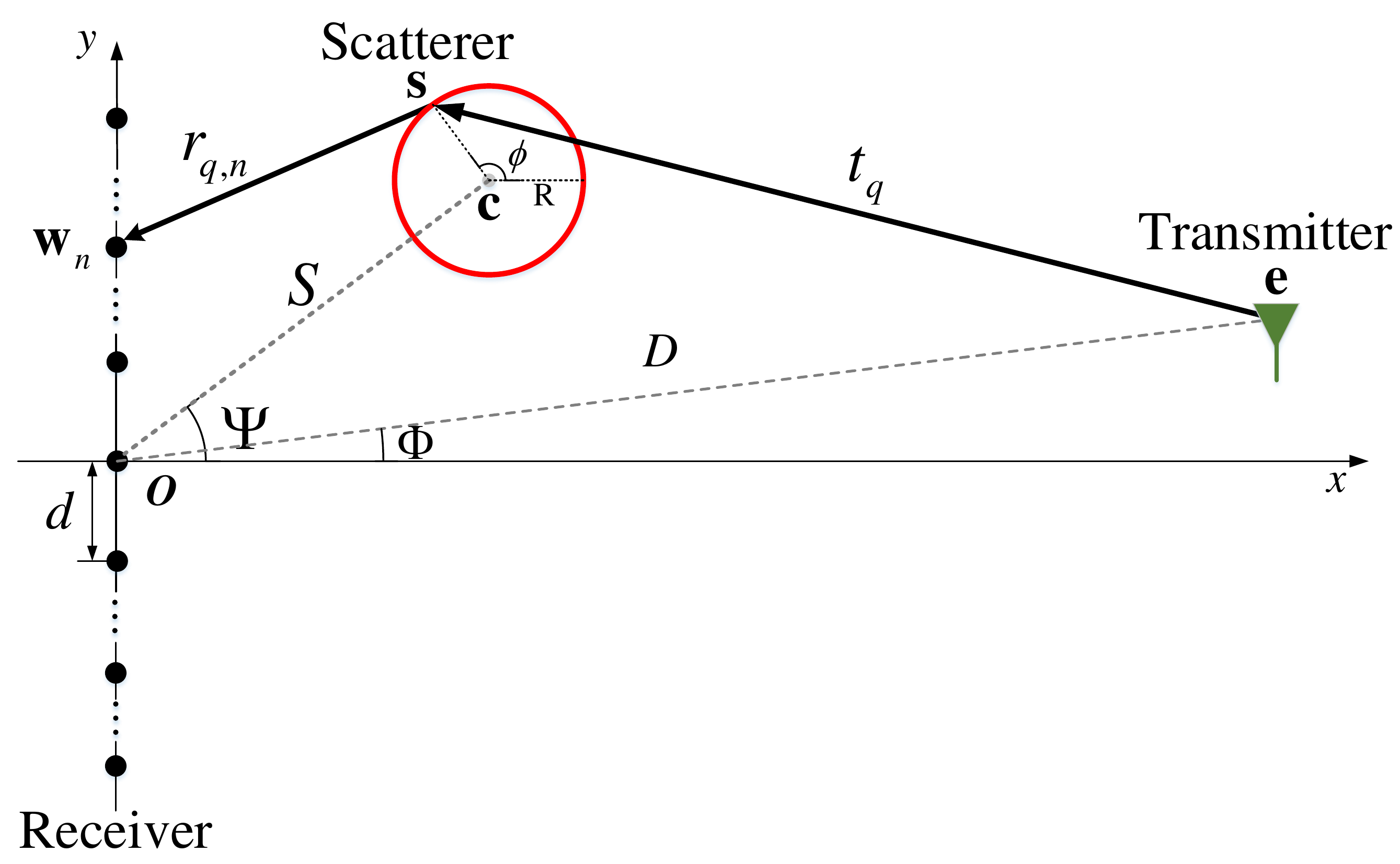}
  \caption{Generalized one-ring model for near-field spatial correlation.}
  \label{fig:Spherical wave2}
\end{figure}
The integral expression in (9) gives a generic near-field spatial correlation for any scatterer distribution $f(\vec{s})$. To gain further insights, in this section, we consider one specific $f(\vec{s})$, termed as generalized one-ring model.
Different from the conventional one-ring model for far-field spatial correlation where the ring center coincides with the center of the antenna array~\cite{837052},
the generalized one-ring model offers more flexibility by allowing the ring center to be freely located, which is more suitable for XL-array communications.
As illustrated in Fig.2, let the radius and ring center be denoted as $R$ and $\vec{c}=[S\cos{\Psi},S\sin{\Psi}]^T$, respectively,
where $S\geq 0$ and $\Psi\in\left(-\frac{\pi}{2},\frac{\pi}{2}\right)$ are the distance and the angle of ring center, respectively.
If $S=0$, the generalized one-ring model reduces to the conventional one-ring model~\cite{837052}.

With generalized one-ring model, the scatterer location can be parameterized as $\vec{s}=\vec{c}+[R\cos{\phi},R\sin{\phi})^T$, where $\phi\in[-\pi,\pi)$ is the angle parameter along the ring.
Therefore, $r_n(\vec{s})$ in (9) is only parameterized by $\phi$, i.e., $r_n(\vec{s})=r_n(\phi)$, which is written as
\begin{equation}\label{eq:r_nS}
r_n(\phi)=\sqrt{(S\cos{\Psi}+R\cos{\phi})^2+(S\sin{\Psi}+R\sin{\phi}-nd)^2}.
\end{equation}

Similarly, we may write $f(\vec{s})=f(\phi)$.
By substituting (\ref{eq:r_nS}) into (\ref{eq:cov1nm}), we have
\begin{equation}\label{eq:cov123}
\begin{aligned}
&\vec{R}_\text{NF}(n,m)=\beta_0\int_{-\pi}^{\pi} \frac{r^2(\phi)e^{-j\frac{2\pi }{\lambda}\left(r_n(\phi)-r_m(\phi)\right) }}{r_n(\phi)r_m(\phi)}f(\phi)\text{d}\phi.
\end{aligned}
\end{equation}

When $Nd\ll r(\phi)$ so that the far-field assumption holds for the generalized one-ring model,
we have $r_n(\phi)\approx r(\phi)-nd(S\sin\Psi+R\sin\phi)/r(\phi)$. In this case, (13) reduces to
\begin{equation}\label{eq:cov_we}
\vec{R}_\text{FF}(n,m)=\beta_0\int_{-\pi}^{\pi}e^{-j\frac{2\pi}{\lambda}(m-n)d\frac{S\sin\Psi+R\sin\phi}{\sqrt{S^2+R^2+2SR\cos(\Psi-\phi)}}}f(\phi)\text{d}\phi.
\end{equation}

To further simplify the expression in (13), we consider the case when $S\gg R$, for which we have the following result.

\emph{Lemma 2:} When $S\gg R$ for the generalized one-ring model, the near-field spatial correlation in (13) can be expressed as
\begin{equation}\label{eq:r_nmApp}
\vec{R}_\text{NF}(n,m)\approx\beta_0\int_{-\pi}^{\pi} \frac{e^{-j\frac{2\pi}{\lambda}\left(S(\sqrt{a_n}-\sqrt{a_m})+R\left(\frac{b_n}{\sqrt{a_n}}-\frac{b_m}{\sqrt{a_m}}\right)\right)}}{\sqrt{a_ma_n}}f(\phi)\text{d}\phi,
\end{equation}
where $a_n=1+\left(nd/S\right)^2-2\left(nd/S\right)\sin{\Psi}$ and $b_n=\cos(\Psi-\phi)-\left(nd/S\right)\sin{\phi}$.
\begin{proof}
Please refer to Appendix B.
\end{proof}

\emph{Lemma 3:} When $S\gg R$ for the generalized one-ring model, the far-field spatial correlation in (\ref{eq:cov_we}) can be expressed as
\begin{equation}\label{eq:cov_we1}
\vec{R}_\text{FF}(n,m)\approx\beta_0\int_{-\pi}^{\pi}e^{-j\frac{2\pi}{\lambda}(m-n)d\left(\sin\Psi+\frac{R}{S}\cos\Psi\sin(\phi-\Psi)\right)}f(\phi)\text{d}\phi.
\end{equation}
\begin{proof}
Please refer to Appendix C.
\end{proof}
To get the closed-form expressions for (15) and (16), we consider the particular von-Mises PDF~\cite{abdi2002parametric}
\begin{equation}\label{eq:Von-Mises}
f(\phi)=e^{\kappa\cos(\phi-\mu)}/(2\pi I_0(\kappa)),  \phi\in[-\pi,\pi),
\end{equation}
where $I_0(\cdot)$ is the zero-order Bessel function of the first kind, $\kappa\geq0$ determines the concentration of the distribution, and $\mu$ corresponds to the angle where the PDF has the peak.
For $\kappa=0$, we have $f(\phi)=\frac{1}{2\pi}$, while for $\kappa=\infty$, $f(\phi)=\delta(\phi-\mu)$.

Define $c_{nm}=\frac{2\pi R}{\lambda} (\frac{1}{\sqrt{a_n}}-\frac{1}{\sqrt{a_m}})$, $d_{nm}=\frac{2\pi Rd}{\lambda S}(\frac{n}{\sqrt{a_n}}-\frac{m}{\sqrt{a_m}})$
and $e_{nm}=\frac{2\pi R(n-m)d\cos{\Psi}}{\lambda S}$.
By substituting (\ref{eq:Von-Mises}) into (\ref{eq:r_nmApp}), the closed-form expression for $\vec{R}_\text{NF}$ is
\begin{equation}\label{eq:NFclose}
\begin{aligned}
&\vec{R}_\text{NF}(n,m)\approx\beta_0\frac{e^{-j\frac{2\pi}{\lambda}S(\sqrt{a_n}-\sqrt{a_m})}}{\sqrt{a_ma_n}I_0(\kappa)}I_0\big(j\{\kappa^2-c_{nm}^2-d_{nm}^2\\
&+2c_{nm}d_{nm}\sin{\Psi}+2j\kappa(d_{nm}\sin(\mu)-c_{nm}\cos(\mu-\Psi)\}^{\frac{1}{2}}\big).\\
\end{aligned}
\end{equation}

Similarly, the closed-form expression for $\vec{R}_\text{FF}$ in (\ref{eq:cov_we1}) with $f(\phi)$ given by (17) can be obtained as
\begin{equation}\label{eq:FFclose}
\begin{aligned}
&\vec{R}_\text{FF}(n,m)\approx\frac{\beta_0e^{-j\frac{2\pi}{\lambda}(m-n)d\sin\Psi}}{I_0(\kappa)}\\
&\times I_0\bigl(j\big\{\kappa^2-e_{nm}^2+2j\kappa e_{nm}\sin(\mu-\Psi)\big\}^{\frac{1}{2}}\bigl).\\
\end{aligned}
\end{equation}

\section{Numerical Results}
\begin{figure}
  \centering
  \includegraphics[width=0.98\columnwidth]{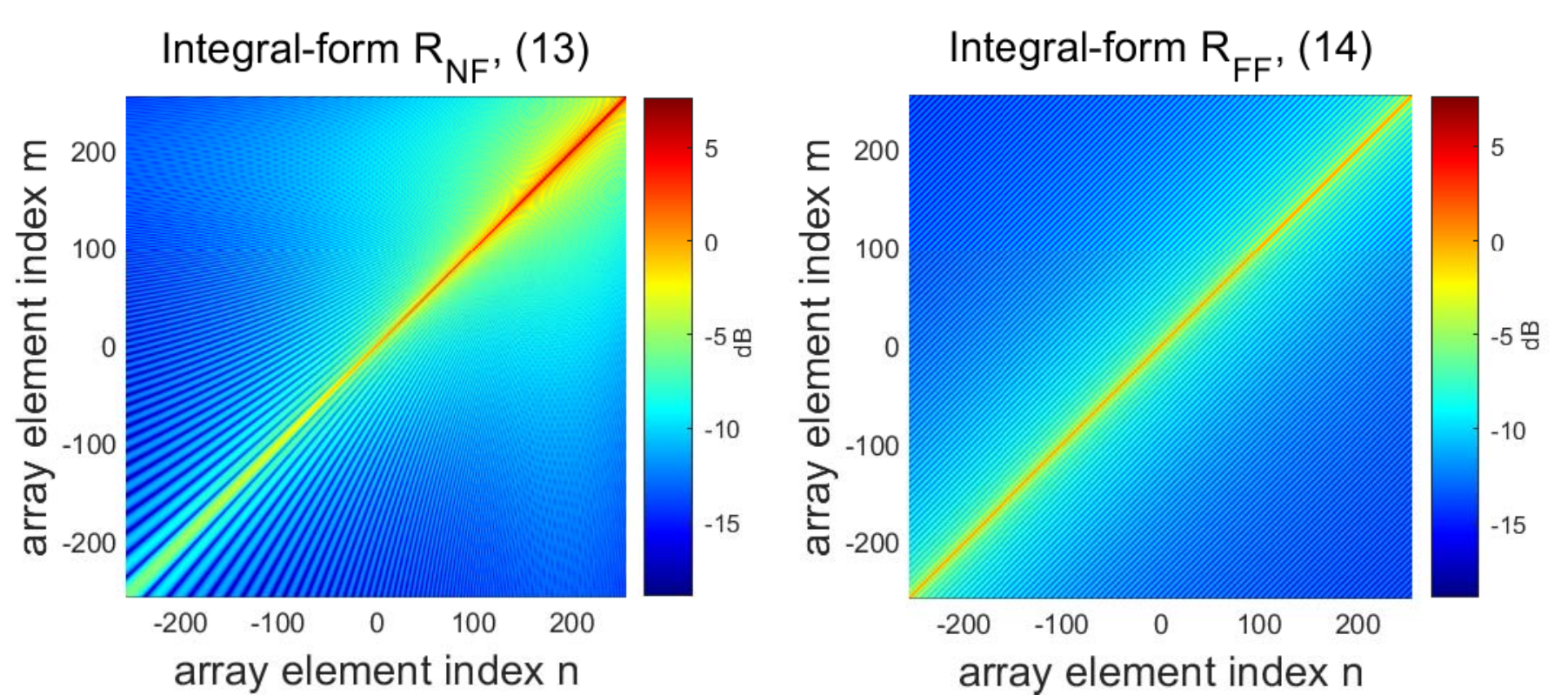}
  \caption{Heatmaps of the near- and far-field spatial correlation matrices.}
  \label{fig:simulation_1}
\end{figure}
In this section, numerical results are provided to compare our developed near-field spatial correlation model with the conventional far-field model.
The number of array elements for the XL-array is $N=512$, with adjacent elements separated by
 $d=\frac{\lambda}{2}$, and the carrier frequency is set as 3.5 GHz.
For the generalized one-ring model, we set $R=3$ m, $\Psi=\frac{\pi}{3}$ and $S$ ranging from $10$ m to $70$ m.
Besides, we set $\kappa=0$ in (17),  i.e., $\phi$ is uniformly distributed in $[-\pi, \pi)$.
Furthermore, the subsequent results are presented by normalizing the spatial correlation by the average power $\beta_0$.


Fig.\ref{fig:simulation_1} shows the heatmaps of the near- and far-field spatial correlation matrices by numerically evaluating the integral-form $\vec{R}_\text{NF}$ and $\vec{R}_\text{FF}$ in (\ref{eq:cov123}) and (\ref{eq:cov_we}), respectively, where $S = 10$ m.
It is observed that the far-field model exhibits a banded pattern,
since $\vec{R}_{\text{FF}}(n,m)$ only depends on the difference $m-n$ due to the spatial wide-sense stationarity.
By contrast, such a spatial stationary property no longer holds for the near-field model,
since $\vec{R}_\text{NF}(n,m)$ depends on indices $n$ and $m$.
Fig.3 also shows that different from the far-field UPW model, the near-field NUSW model leads to non-uniform average power across array elements,
which is expected due to the distance variations from each scatterer to different array elements.

\begin{figure}
  \centering
 \includegraphics[width=0.85\columnwidth]{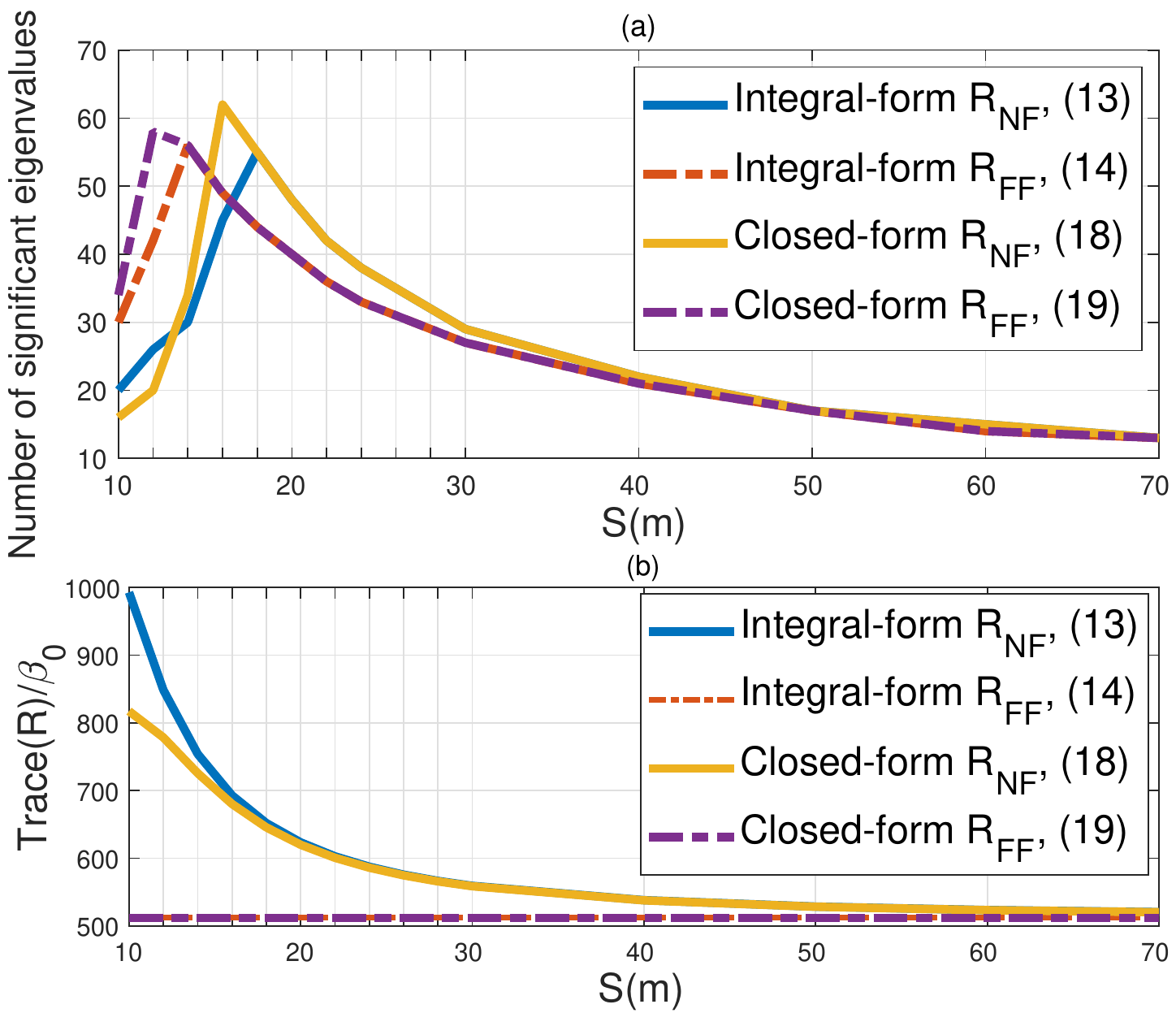}
  \caption{(a) The number of significant eigenvalues for the near- and far-field spatial correlations;
  (b) $\text{Trace}(\vec{R})/\beta_0$ for the near- and far-field spatial correlations.}
  \label{fig:simulation_4}
\end{figure}

Fig.4 compares the number of ``significant" eigenvalues as well as the total normalized power $\text{trace}(\vec{R})/\beta_0$ of the near- and
far-field spatial correlation matrices versus the ring center distance $S$, where an eigenvalue is regarded as ``significannt'' when it is no smaller than $1\%$ of $\text{trace}(\vec{R})/\beta_0$.
Both the integral forms in (13), (14), and the closed forms in (18), (19) are evaluated.
It is observed from Fig.4(a) that when $S$ is small, the near- and far-field models lead to significantly different results. For example,
when $S=14$ m, the number of significant eigenvalues of the far-field model is about twice of that of the near-field spatial correlation matrix.
Furthermore, it is also observed from Fig.\ref{fig:simulation_4}(b) that the total normalized channel power $\text{trace}(\vec{R}_\text{FF})/\beta_0=512$, which is independent of $S$. By contrast, that for
 the near-field model decreases as $S$ increases. Such differences demonstrate the importance of proper near-field modelling for XL-array communications.
\begin{figure}
  \centering
 \includegraphics[width=0.85\columnwidth]{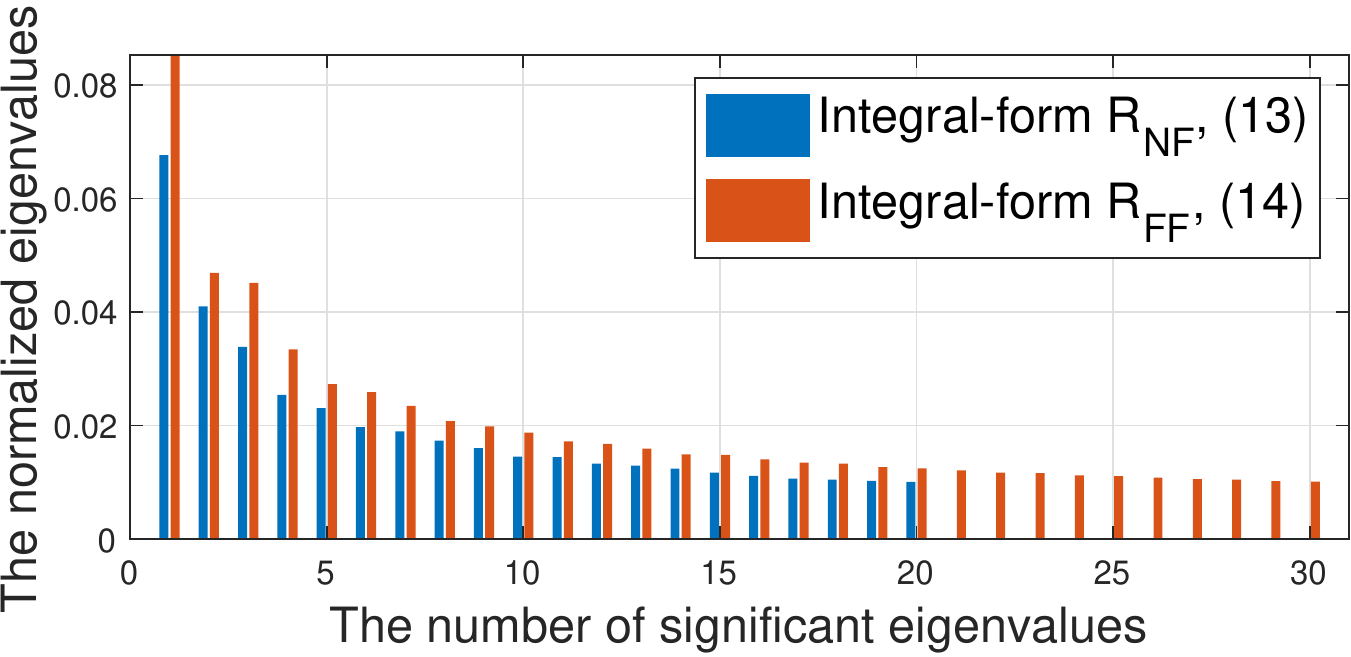}
  \caption{The normalized eigenvalue distributions for the near- and far-field spatial
correlations, when $S=10$ m.}
  \label{fig:simulation_3}
\end{figure}
\begin{figure}
  \centering
  \includegraphics[width=0.85\columnwidth]{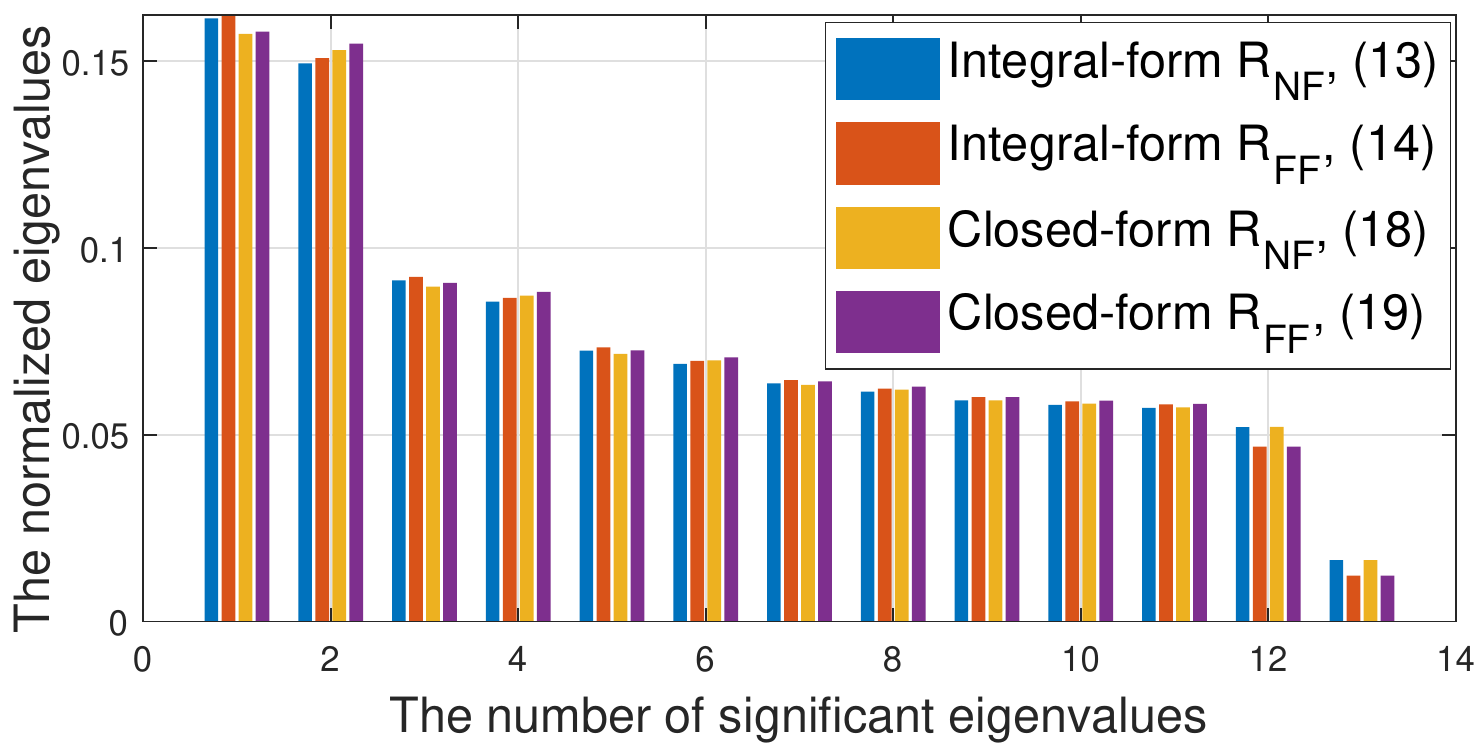}
  \caption{The normalized eigenvalue distributions for the near- and far-field spatial correlations, when $S=70$ m.}
  \label{fig:simulation_6}
\end{figure}

Finally, Fig.5 and Fig.6 plot the eigenvalue distributions of the correlation matrices for $S=10$ m and $S=70$ m, respectively.
For $S=10$ m, the result is only evaluated by the integral (13) and (14), since the assumption of $S\gg R$ for the closed-form expressions (18) and (19) does not hold.
It is observed from Fig.5 that for $S=10$ m, the near- and far-field models lead to significantly different eigenvalue distributions.
By contrast, when $S=70$ m, 
both the near- and far-field models lead to roughly the same eigenvalue distributions.
This demonstrates that the developed near-field spatial correlation is more generic than the far-field model by including it as a special case when the scatterers are far from the XL-array.

\section{Conclusion}
This letter studied the channel modelling and spatial correlation characterization for XL-array communications, by taking into account the NUSW characteristics.
A novel integral expression was derived for the near-field spatial correlation, which generalized the conventional far-field spatial correlation.
The proposed near-field spatial correlation depends
on the PLS, which is different from the far-field spatial correlation that only depends on the PAS.
Furthermore, it is revealed that the near-field spatial correlation was no longer spatially wide-sense stationary as in the far-field modelling.
Simulation results demonstrated the necessity of the near-field spatial correlation characterization for XL-array communications.

\begin{appendices}
 \section{ PROOF OF LEMMA 1}
When $Nd\ll r(\vec{s})$, $r_n(\vec{s})$ can be approximated as
\begin{equation}\label{eq:r_n}
r_n(\vec{s})\approx r(\vec{s})\sqrt{1-2nd\sin(\theta(\vec{s}))/r(\vec{s})}\overset{a}{\approx}r(\vec{s})-nd\sin(\theta(\vec{s})),
\end{equation}
where $\theta(\vec{s})$ represents the angle of scatterer $\vec{s}$ with respect to the positive $x$-axis and $(a)$ follows from the first-order Taylor approximation $\sqrt{1+x}\approx1+\frac{1}{2}x$ for $x\rightarrow 0$.
By substituting (\ref{eq:r_n}) into (\ref{eq:cov1nm}), the integral function is only related to $\theta(\vec{s})$. Hence, (\ref{eq:cov1nm}) reduces to
%
\begin{equation}\label{eq:cov3}
\vec{R}_\text{NF}(n,m)\approx\beta_0\int_{-\pi}^{\pi} e^{-j\frac{2\pi }{\lambda}(m-n)d\sin(\theta)} f(\theta)\text{d}\theta.\\
\end{equation}

This thus completes the proof of Lemma 1.
 \section{PROOF OF LEMMA 2}
When $S\gg R$, we have $r(\vec{s})\approx S$, $\forall\vec{s}\in\vec{S}$. Hence, $r_n(\vec{s})$ can be approximated as
\begin{equation}\label{eq:r_app1}
\begin{aligned}
r_n(\vec{s})\approx S\sqrt{a_n+2b_nR/S},\\
\end{aligned}
\end{equation}
where $a_n=1+\left(nd/S\right)^2-2nd\sin{\Psi}/S$ and $b_n=\cos(\Psi-\phi)-\left(nd/S\right)\sin{\phi}$.
Based on the definition of $a_n$, 
we have $a_n>0$.
Therefore, (\ref{eq:r_app1}) is simplified to
\begin{equation}\label{eq:r_app2}
\begin{aligned}
r_n(\vec{s})=S\sqrt{a_n}\sqrt{1+2(b_n/a_n)(R/S)}.
\end{aligned}
\end{equation}

Next, 
we make a further transformation of $b_n$, i.e., $b_n=\sqrt{a_n}\sin(\phi+\vartheta)$, where $\sin(\vartheta)=\frac{\cos{\Psi}}{\sqrt{a_n}}$ and $\cos(\vartheta)=\frac{\sin{\Psi}-nd/S}{\sqrt{a_n}}$.
Hence, $\frac{b_n}{a_n}=\frac{\sin(\phi+\vartheta)}{\sqrt{a_n}}$. Since $\sin(\phi+\vartheta)\in[-1,1]$ and $\sqrt{a_n}>0$, the value of $\frac{b_n}{a_n}$ is finite.
As a result, $r_n(\vec{s})$ can be approximated as
\begin{equation}\label{eq:r_app3}
r_n(\vec{s})\overset{a}{\approx} S\sqrt{a_n}+Rb_n/\sqrt{a_n}.
\end{equation}

By applying $r_n(\vec{s})$ in (\ref{eq:r_app3}) into (\ref{eq:cov123}), $\vec{R}_\text{NF}$ is simplified as
\begin{equation}\label{eq:r_nmApp1}
\begin{aligned}
\vec{R}_\text{NF}(n,m)\approx\beta_0\int_{-\pi}^{\pi} &\frac{e^{-j\frac{2\pi}{\lambda}\left(S(\sqrt{a_n}-\sqrt{a_m})+R\left(\frac{b_n}{\sqrt{a_n}}-\frac{b_m}{\sqrt{a_m}}\right)\right)}}{\sqrt{a_ma_n}}f(\phi)\text{d}\phi.\\
\end{aligned}
\end{equation}

The proof of Lemma 2 is thus completed.

\section{PROOF OF LEMMA 3}
When $S\gg R$, the component $\frac{S\sin\Psi+R\sin\phi}{\sqrt{S^2+R^2+2SR\cos(\Psi-\phi)}}=c$ in (\ref{eq:cov_we}) can be approximated as
\begin{equation}\label{eq:ed}
c\overset{a}{\approx} \frac{\sin\Psi+\frac{R}{S}\sin\phi}{1+\frac{R}{S}\cos(\Psi-\phi)}\approx \sin\Psi+\frac{R}{S}\cos\Psi\sin(\phi-\Psi).
\end{equation}

By  substituting (\ref{eq:ed}) into (\ref{eq:cov_we}), $\vec{R}_\text{FF}$ in (\ref{eq:cov_we}) is simplified as
\begin{equation}\label{eq:cov_we3}
\vec{R}_\text{FF}(n,m)\approx\beta_0\int_{-\pi}^{\pi}e^{-j\frac{2\pi}{\lambda}(m-n)d\left(\sin\Psi+\frac{R}{S}\cos\Psi\sin(\phi-\Psi)\right)}f(\phi)\text{d}\phi.
\end{equation}

The proof of Lemma 3 is thus completed.
\end{appendices}

\bibliographystyle{IEEEtran}
\bibliography{\bibdir/header_short,\bibdir/bibliography}
\end{document}

%% file: Definitions.tex
\def\mb{\mathbf}

\def\rm{\textrm}

\renewcommand{\vec}{\mb}    